\documentclass[twocolumn,superscriptaddress,aps,prl]{revtex4-1}
\usepackage{amsmath}
\usepackage{graphicx}
\usepackage{dcolumn}
\usepackage{color}
\usepackage{ulem}

\usepackage{epstopdf}
\usepackage{color}
\usepackage{subfigure}
\usepackage{float}
\usepackage{mathrsfs}
\usepackage{bbold}
\usepackage{psfrag}
\usepackage{mathcomp}
\usepackage{verbatim}
\usepackage{multirow}
\usepackage{diagbox}
\usepackage[colorlinks,citecolor=blue]{hyperref}

\usepackage{simplewick}

\begin{document}

\title{Non-Hermitian Linear Response Theory}
\author{Lei Pan}
\affiliation{Institute for Advanced Study, Tsinghua University, Beijing,
100084, China}
\author{Xin Chen}
\affiliation{Institute for Advanced Study, Tsinghua University, Beijing,
100084, China}
\author{Yu Chen}
\email{spaceexplorer@163.com}
\affiliation{Center for Theoretical Physics and Department of Physics, Capital Normal University, Beijing, 100048, China}
\author{Hui Zhai}
\email{hzhai@tsinghua.edu.cn}
\affiliation{Institute for Advanced Study, Tsinghua University, Beijing, 100084, China}

\begin{abstract}
Linear response theory lies at the heart of studying quantum matters, because it connects the dynamical response of a quantum system to an external probe and correlation functions of the unprobed equilibrium state. Thanks to the linear response theory, various experimental probes can be used for determining equilibrium properties. However, so far both the unprobed system and the probe operator are limited to Hermitian ones. Here we develop a non-Hermitian linear response theory that considers the dynamical response of a Hermitian system to a non-Hermitian probe, and we can also relate such dynamical response to properties of unprobed Hermitian system at equilibrium. As an application of our theory, we consider the real-time dynamics of momentum distribution induced by one-body and two-body dissipations. Remarkably,  we find that, for critical state with no well-defined quasi-particles, the dynamics are slower than normal state with well-defined quasi-particles, and our theory provides a model independent way to extract the critical exponent in the real time spectral function. For demonstration, we find surprising good agreement between our theory and a recent cold atom experiment on the dissipative Bose-Hubbard model. We also propose to further quantitatively verify our theory by performing experiments on dissipative one-dimensional Luttinger liquid. 

\end{abstract}

\maketitle

Most experiments on condensed matter and cold atom systems measure how an observable changes after perturbing the system. The linear response theory is crucial for the interpretation of these experiments because it attributes such dynamical responses to Green functions of unperturbed system at equilibrium, under the condition that the perturbation does not bring the system far away from equilibrium \cite{LinearResponse}. Because of the linear response theory, these experimental measurements can be used to reveal equilibrium properties of quantum matters. The linear response theory has broad applications for experiments ranging from conductivity measurements, angle-resolved photoemission spectroscopy and neutron scattering in condensed matter physics to the radio-frequency and lattice modulation spectroscopies in cold atom physics. 

Here, we consider applying a non-Hermitian perturbation to a Hermitian quantum system, and then monitoring how an  observable changes due to this non-Hermitian perturbation. With the same spirit of the linear response theory, we work out the general relation that can also relate this dynamical response to properties of the unperturbed Hermitian system at equilibrium. Therefore, we name our theory as \textit{the non-Hermitian linear response theory}. Recently, there are growing number of experiments on quantum systems with tunable dissipations \cite{BHMExp,Takahashi1,Takahashi2,Luo,Cirac,Barontini}, and our theory is perfectly suitable for analyzing these experiments which are performed in weak dissipation regime. And as we will show in an example below, such analysis can indeed yield new insight for these non-Hermitian experiments.  

We would like to emphasize the difference between this work and the existing works. On the linear response theory side, previous works only consider the situations that the perturbation is a Hermitian one. We compare our results with the textbook results on Hermitian perturbation in Table \ref{comparison}. On the non-Hermitian physics and dissipation physics side, although recently there are many works on non-Hermitian and dissipative systems \cite{Lee16,Nori17,Fu18,Wang18,Bergholtz18,Ueda18,Zhai18,Ueda18b,Tom19,Ashvin19,Szameit15,Fu18ex,TopLaserT,TopLaserE,Tom19ex,Xue19ex}, they all consider the Hermitian and non-Hermitian parts together as a whole, and these works are interested in how the non-Hermitian part can drive new interesting physics that does not exist for Hermitian part alone. They do not implement the idea of linear response to treat the non-Hermitian part as a detection tool in order for revealing equilibrium property of the Hermitian part. Comparing to the common approach of solving the full Lindblad equation, we consider the situation that the dissipation is weak such that dissipation time is longer comparing to the intrinsic relaxation time of the system. Hence we can take the ensemble average of the perturbation operator under unperturbed equilibrium state. This is why we can obtain simplified and universal results and extract the important information of equilibrium state from the dissipative dynamics.       

As an application of our non-Hermitian linear response theory, we determine how the real time dynamics of the momentum distribution responding to one-body and two-body dissipations. We show that this response is determined by a universal function given by the single-particle real time spectral function at equilibrium. We consider two different kinds of quantum states, one being normal state with well-defined quasi-particle and the other being critical state without quasi-particles. These two types of systems have different single-particle spectral functions, and as a result, we show that the dynamics for critical states is generally slower than that for normal states, given the same strength of dissipations. For critical states such as the Luttinger liquid, non-Fermi liquids or states in quantum critical regimes \cite{Sachdev}, we show that the critical exponent in the spectral function can be extracted from the dissipation induced dynamics of the momentum distribution. 

One striking prediction from our theory is that, in many cases, the decay of the height and the broadening of the width of the momentum distribution are governed by the universal function.  We show that this predication can be successfully confirmed by reanalyzing experimental data on dissipative two-dimensional Bose-Hubbard model \cite{BHMExp}, which strongly supports our theory. We propose to further benchmark our theory in dissipative Luttinger liquids, where the critical exponents measured by our method can be quantitatively compared with theory. Then, our protocol can be applied to cases like two-dimensional superfluid-Mott insulator transition, where the critical exponent is hard to calculate in theory. 

\begin{widetext}
\hspace{3ex}\begin{table}
\begin{tabular}{|c|c|c|}
\hline
&Form of Perturbation &Response Theory\\
\hline
Hermitian&$\gamma \sum_j\hat{\cal O}^\dag_j\hat{\cal O}_j$&$\delta\mathcal{W}(t)=-i\gamma\int_0^t \sum_j\langle[\hat{W}(t),\hat{\cal O}^\dag_j(t')\hat{\cal O}_j(t')]\rangle dt'$\\
\hline
non-Hermtian&$-i\gamma \sum_j\hat{\cal O}^\dag_j\hat{\cal O}_j+\sum_j\hat{\cal O}^\dag_j\xi_j+\xi^\dag_j\hat{\cal O}_j$ &$\delta \mathcal{W}(t)=-\gamma\int_0^t\sum_j\langle\{\hat{W}(t),\hat{\cal O}_j^\dag(t')\hat{\cal O}_j(t')\}-2\hat{\cal O}_j^\dag(t')\hat{W}(t)\hat{\cal O}_j^{}(t')\rangle dt'$\\
\hline
\end{tabular}
\caption{Comparison of the formalism between the Hermitian linear response theory and the non-Hermtian linear response theory. The system itself is a Hermitian one, but the perturbation can either be a Hermitian one or a non-Hermitian one. Response of observable $\mathcal{W}$ to these two different types of perturbations are compared.   \label{comparison} }
\end{table}
\end{widetext}

\textbf{General Formalism.} 

Let us consider a Hermitian system with Hamiltonian $\hat{H}_0$ and a non-Hermitian dissipation term $\hat{H}_{\rm diss}$ added as perturbation. Here $\hat{H}_{\text{diss}}$ contains both a non-Hermitian dissipation term and a corresponding Langevin noise term as 
 \begin{eqnarray}
\hat{H}_{\rm diss}=\sum_j\left(-i\gamma \hat{\cal O}_j^\dag \hat{\cal O}_j^{}+\hat{\cal O}_j^\dag\xi_j^{}+\xi^\dag_j \hat{\cal O}_j^{}\right).
\end{eqnarray}
Here without loss of generality, we introduce a Hermitian operator as $\sum_j\hat{\cal O}_j^\dag \hat{\cal O}_j^{}$ and its coefficient is purely imaginary ($\gamma$ is real), which gives a non-Hermitian perturbation. $\xi$ and $\xi^\dag$ are the Langevin noise operators, and they satisfy $\langle \hat{\xi}_{j}(t)\hat{\xi}_{\ell}^\dag(t')\rangle_{\rm noise}=2\gamma\delta_{j\ell}\delta(t-t')$, $\langle \hat{\xi}_j^\dag(t)\hat{\xi}^\dag_l(t')\rangle_{\rm noise}=0$, where $\langle\cdot\rangle_{\rm noise}$ stands for averaging over noise configurations \cite{Kamenev}. The Langevin noise terms always come together with the non-Hermitian dissipation term if the dissipation arises from the coupling the Hermitian system to a large environment, and the presence of the Langevin force is important to ensure the unitary of quantum operators \cite{Kamenev}.

Let us consider a Hermitian operator $\hat{W}$ and the physical observable $\mathcal{W}$ is given by
\begin{equation}
\mathcal{W}=\langle \text{Tr}(\rho_0\hat{W}_\text{H}(t))\rangle_{\rm noise}.
\end{equation} 
Here $\rho_0=e^{-\beta\hat{H}_0}/\mathcal{Z}$ is the initial equilibrium density matrix of the non-perturbed Hermitian system with temperature $T=1/\beta$, where $\mathcal{Z}=\text{Tr}(e^{-\beta\hat{H}_0})$ is the partition function of the system. The trace of the density matrix $\rho_0$ is preserved with the help of the Langevin noise term. Also because of the Langevin noise term, there is an extra noise average in addition to the usual ensemble average. $\hat{W}(t)$ is the operator in the Heisenberg picture given by 
\begin{eqnarray}
\hat{W}_H(t)=e^{i\hat{H}^\dag t}\hat{W} e^{-i\hat{H}t},\label{Eq:Lindbald}
\end{eqnarray}
where $\hat{H}=\hat{H}_0+\hat{H}_{\rm diss}$. 

The derivation of the linear response follows straightforwardly from the perturbation expansion in term of $\hat{H}_\text{diss}$. However, we should note a major difference comparing to the Hermitian linear response theory. In the Hermitian case, the first order perturbation gives the contribution to the linear order of coupling constant $\gamma$. Here we would also like to keep the linear order contribution of $\gamma$, but this requires us to keep the first order perturbation of the dissipation term and the second order of the Langevin noise term, because, as mentioned above, the the second order of $\xi$ is proportional to $\gamma$. A straightforward calculation yields \cite{supple}
\begin{align}
&\delta\mathcal{W}(t)\equiv\mathcal{W}(t)-\mathcal{W}(0)\nonumber\\
&=-\gamma\sum_j\!\!\int_0^t\!\!\langle\{\hat{W}(t),\hat{\cal O}_j^\dag(t^\prime)\hat{\cal O}_j(t') \}\!-\! 2\hat{\cal O}_j^\dag(t')\hat{W}(t)\hat{\cal O}_j(t')\rangle dt',\label{formula}
\end{align}
where $\{\dots,\dots\}$ denotes the anti-commutator. The ensemble average is taken over the unperturbed equilibrium state. Here the first term of Eq. \ref{formula} follows from the first order perturbation in the dissipation term and it is similar as the Hermitian case (see Table \ref{comparison} for comparison) except the commutator is replaced by the anti-commutator. The second term of Eq. \ref{formula} has no counter-part in the Hermitian case, which arises from the second order perturbation in term of the Langevin noise term. In Table \ref{comparison}, we compare how the observable $\mathcal{W}$ responds to a Hermitian perturbation and to a non-Hermitian perturbation. In both cases, the responses are attributed to correlations at unperturbed equilibrium state.

\textbf{Application to Momentum Distribution.} 

Now we apply our theory to discuss how momentum distribution responds to one-body and two-body dissipations, which are typical experimental measurements in cold atom systems. As an example, we consider a lattice boson system, with $\hat{a}_j$ being the boson creation operator at site-$j$. The discussion can be directly applied to a continuous system by replacing $\hat{a}_j$ with $\hat{a}({\bf r})$ and replacing summation over $j$ with integration over space. The results can also be straightforwardly generalized to fermions too.  Here we take $\hat{A}=\hat{n}_{{\bf k}}=\hat{a}^\dag_{{\bf k}}\hat{a}_{{\bf k}}$, $\hat{O}_j=\hat{a}_j$ for one-body dissipation and $\hat{O}_j=\hat{n}_j$ for two-body dissipation. 

\textit{One-Body Dissipation.} With our formula Eq. \ref{formula}, for one-body dissipation it yields
\begin{eqnarray}
\delta n_{\bf k}(t)=&&\gamma \sum_i\left[\int_0^t dt' 2\langle \hat{a}_i^\dag(t')\hat{a}^\dag_{\bf k}(t) \hat{a}_{\bf k}^{}(t) \hat{a}_i^{}(t')\rangle-\right.\nonumber\\
&&\left.\int_0^t dt' \langle  \{\hat{a}_i^\dag(t')\hat{a}_i^{}(t'),\hat{a}^\dag_{\bf k}(t) \hat{a}_{\bf k}^{}(t)\}\rangle\right].\label{nk_one}
\end{eqnarray}
The r.h.s of Eq. \ref{nk_one} are four-point correlation functions at equilibrium. With the temporal translational symmetry, we define $G_{\bf k}^<(t)=-i\langle a^\dag_{\bf k}(t)a^{}_{\bf k}(0)\rangle$, and we introduce the spectral function $\mathcal{A}_{\bf k}(\omega)$ via $G_{\bf k}^{<}(t)=\int d\omega \mathcal{A}_{\bf k}(\omega)n_\text{B}(\omega)e^{i\omega t}$, where $n_\text{B}(\omega)$ is the Bose distribution function. By applying the generalized Wick's theorem for the Keldysh contour ordered operator product, we can decompose all four-point correlation functions into two-point Green's functions \cite{Pourfath}. With straightforward calculations \cite{supple}, it finally yields a rather simple results that is
\begin{equation}
\delta n_{{\bf k}}(t)=-2\gamma\left(\int_{0}^{t}g({\bf k}, t^\prime)g({\bf k}, -t^\prime)dt^\prime\right)n_{{\bf k}}(0), \label{nk_one_2}
\end{equation} 
where $n_{{\bf k}}(0)$ denotes the initial momentum distribution before turning on the dissipation. Here  
\begin{equation}
g({\bf k},t)\equiv\int d\omega e^{i\omega t}{\cal A}_{\bf k}(\omega),
\end{equation}
and we denote $f({\bf k},t)\equiv g({\bf k}, t)g({\bf k},-t)$ and $\mathcal{F}({\bf k}, t)=\int_0^{t}f({\bf k},t^\prime)dt^\prime$. It turns out that these two functions are the key for following discussions. It is worth noting that both the dissipation term and the Langevin noise term contribute equally to the final results, in particular, some cancellation from two contributions results in such a simple expression. 

\begin{figure}[t]
	\includegraphics[width=.85\columnwidth]{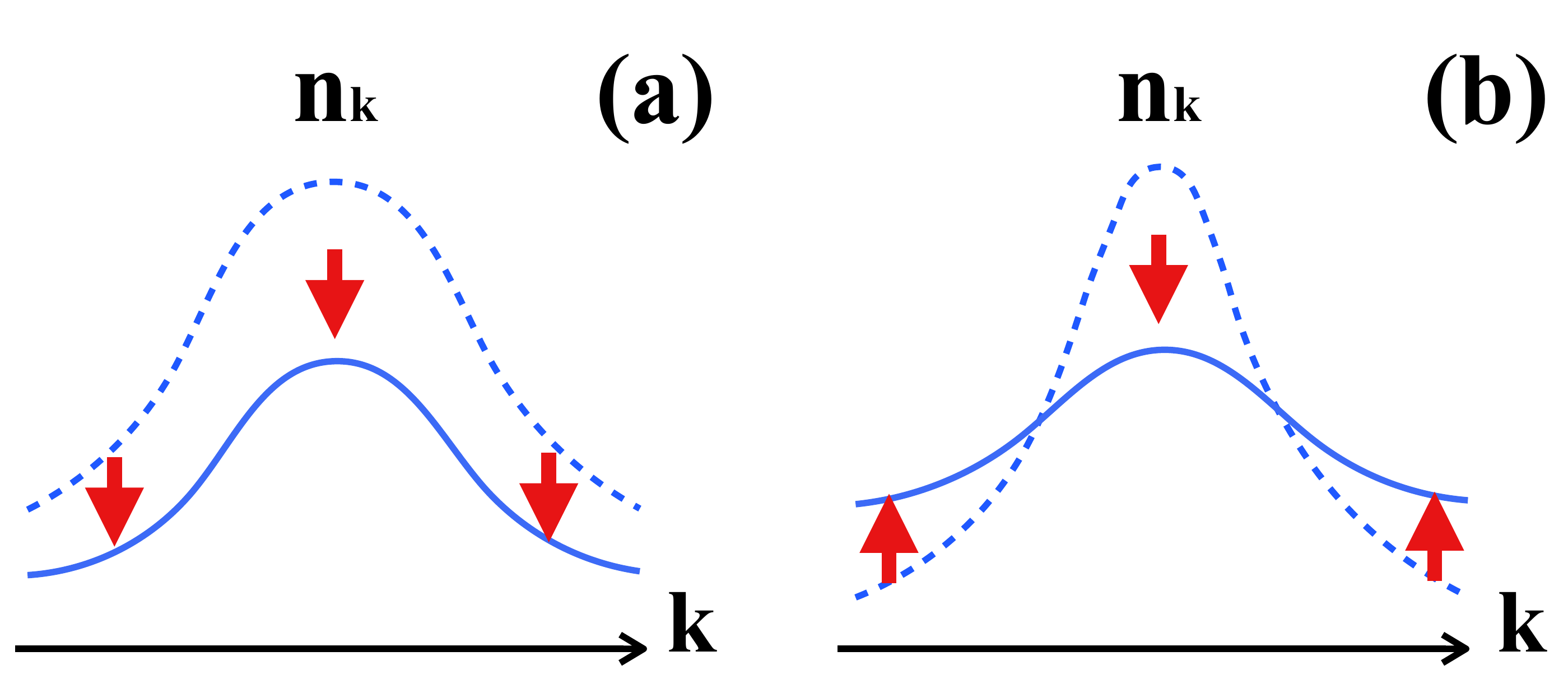}
	\caption{Schematic of real time dynamics of the momentum distribution induced dissipations. (a) one-body dissipation and (b) two-body dissipations. The dashed lines and solid lines denote the momentum distributions at the initial and the later time. Downward and upward arrows indicate decreasing and increasing population, respectively. } 
	\label{One_Two_Body_Loss}
\end{figure}

Under weak dissipation, we consider the entire dynamical process as a quasi-static one, Eq. \ref{nk_one_2} can be written as a differential equation as
\begin{equation}
\frac{d n_{{\bf k}}(t)}{dt}=-2\gamma f({\bf k}, t) n_{{\bf k}}, \label{nk_one_3}
\end{equation}
whose solution gives 
\begin{equation}
n_{{\bf k}}(t)=e^{-2\gamma \mathcal{F}({\bf k},t)}n_{{\bf k}}(0). \label{sol_one}
\end{equation}
Because $g({\bf k},t)=g^*({\bf k}, -t)$, $f({\bf k}, t)$ and $\mathcal{F}({\bf k}, t)$ are both positive. Thus, this result shows that a one-body dissipation leads to decay of populations at all momentum, as shown in Fig. \ref{One_Two_Body_Loss}(a).

\textit{Two-Body Dissipation.} We can also apply Eq. \ref{formula} to the two-body dissipation, and the response is given by 
\begin{eqnarray}
\delta n_{\bf k}(t)=&&\gamma \sum_i\left[\int_0^t dt' 2\langle \hat{n}_i^\dag(t')\hat{a}^\dag_{\bf k}(t) \hat{a}_{\bf k}^{}(t) \hat{n}_i^{}(t')\rangle-\right.\nonumber\\
&&\left.\int_0^t dt' \langle \{\hat{n}_i^\dag(t')\hat{n}_i^{}(t'),\hat{a}^\dag_{\bf k}(t) \hat{a}_{\bf k}^{}(t)\}\rangle\right].\label{nk_two}
\end{eqnarray}
The r.h.s. of Eq. \ref{nk_two} are six-point correlation functions in term of $\hat{a}$-operator. Similarly, by applying the Wick's theorem and casting into differential equations \cite{supple}, it also yields a surprisingly simple form
\begin{equation}
\frac{d\Delta n_{{\bf k}}}{dt}=-2\gamma f({\bf k},t)\Delta n_{{\bf k}}. \label{nk_two_1}
\end{equation} 
This equation is very similar to Eq. \ref{nk_one_3}, and it is quite striking that two dynamics are governed by the same function $f({\bf k}, t)$. The only difference between these two equations is that $n_{{\bf k}}$ in Eq. \ref{nk_one_3} is replaced by  $\Delta n_{{\bf k}}$, which is defined as $n_{{\bf k}}(t)-\bar{n}$ and $\bar{n}$ is the mean density. However, the physical consequences are quite different due to this difference. The solution of Eq. \ref{nk_two_1} is 
\begin{equation}
\Delta n_{{\bf k}}(t)=e^{-2\gamma \mathcal{F}({\bf k},t)}\Delta n_{{\bf k}}(0), \label{sol_two}
\end{equation}
which means that the population decreases for momenta with initially $n_{{\bf k}}(0)>\bar{n}$, and increases for momenta with initially $n_{{\bf k}}(0)<\bar{n}$. As a result, the two-body dissipation broadens the momentum distribution, and eventually, $n_{{\bf k}}$ approaches $\bar{n}$ for all ${\bf k}$, as shown in Fig. \ref{One_Two_Body_Loss}(b).

We note that Eq. \ref{sol_one} and Eq. \ref{sol_two} can be respectively casted into Eq. \ref{sol_one_2} and Eq. \ref{sol_two_2}
\begin{align}
&n_{{\bf k}}(t)-n_{{\bf k}}(0)=-(1-e^{-2\gamma \mathcal{F}({\bf k},t)})n_{{\bf k}}(0). \label{sol_one_2}\\
&n_{{\bf k}}(t)-n_{{\bf k}}(0)=-(1-e^{-2\gamma \mathcal{F}({\bf k},t)})\Delta n_{{\bf k}}(0). \label{sol_two_2}
\end{align}
Eq. \ref{sol_one_2} and Eq. \ref{sol_two_2} show that, given the same dissipation strength, these two time dependences are governed by exactly the same function. Thus, this function $\mathcal{F}({\bf k},t)$ is most crucial quantity that fixes the entire dynamics of momentum distribution. 

\textit{Normal Phase with Well-Defined Quasi-Particles.} Below we consider $\mathcal{F}({\bf k},t)$ for two types of different quantum phases. The first has well defined quasi-particles, whose spectrum functions typically take the form
\begin{equation}
\mathcal{A}(\omega)=\frac{\Gamma_{{\bf k}}}{(\omega-\epsilon_{{\bf k}})^2+\Gamma^2_{{\bf k}}},
\end{equation}
where $1/\Gamma_{{\bf k}}$ stands for the quasi-particle lifetime. In this case, we find that
\begin{equation}
f({\bf k}, t)=e^{-2t\Gamma_{{\bf k}}}, \    \  \mathcal{F}({\bf k}, t)=\frac{1-e^{-2t\Gamma_{{\bf k}}}}{2\Gamma_{{\bf k}}}.
 \end{equation}
When the quasi-particle lifetimes are not sensitive to momentum ${\bf k}$, $f({\bf k},t)$ becomes momentum independent. In particular, when the quasi-particle lifetimes are sufficiently longer than $1/\gamma$, we can take the approximation $\Gamma_{{\bf k}}\rightarrow 0$, and this is equivalent to taking $\mathcal{A}(\omega)=\delta(\omega-\epsilon_{{\bf k}})$. It results in $\mathcal{F}({\bf k},t)=t$, which gives the conventional exponential behavior for the time dynamics. In the small $t$ regime, it gives rise to diffusion dynamics. That is to say, this diffusion dynamics is a common behavior of systems with long-lived quasi-particles. 

\textit{Critical Phase with no Well-Defined Quasi-Particles.} Then we move to consider critical phases with no well-defined quasi-particles. In this case, we have \cite{Zinn-Justin}
\begin{equation}
\mathcal{A}(\omega)\propto \frac{\Theta(\omega-\epsilon_{{\bf k}})}{(\omega-\epsilon_{{\bf k}})^\eta}, \label{spectral_Luttinger}
\end{equation} 
where $\Theta(x)=1$ for $x>0$ and $\Theta(x)=0$ for $x<0$. Luttinger liquids, non-Fermi liquids and low-temperature quantum critical regimes all exhibit such spectral functions. Here $\eta$ is taken as a critical exponent with no ${\bf k}$-dependence. In this case, we obtain
\begin{equation}
f({\bf k}, t)\propto t^{2\eta-2}, \    \ \mathcal{F}({\bf k}, t) \propto t^{2\eta-1},
\end{equation}
where the proportional coefficients are non-universal constants. Then the momentum distribution is determined by a stretched exponential function
\begin{equation}
1-e^{-2\gamma \mathcal{F}({\bf k},t)}=1-e^{-2\kappa t^{2\eta-1}}, \label{critical}
\end{equation}
where $\kappa$ is a constant proportional to $\gamma$. For later connivence, we introduce $\kappa=\tau^{1-2\eta}_0$ which defines a characteristic time scale $\tau_0$. In a physical system, $\eta$ has to be smaller than unity, because of the requirement that the spectral function is normalizable. Therefore, the dynamics is slower comparing to the phases with quasi-particles.

\begin{figure}[t]
	\includegraphics[width=.9\columnwidth]{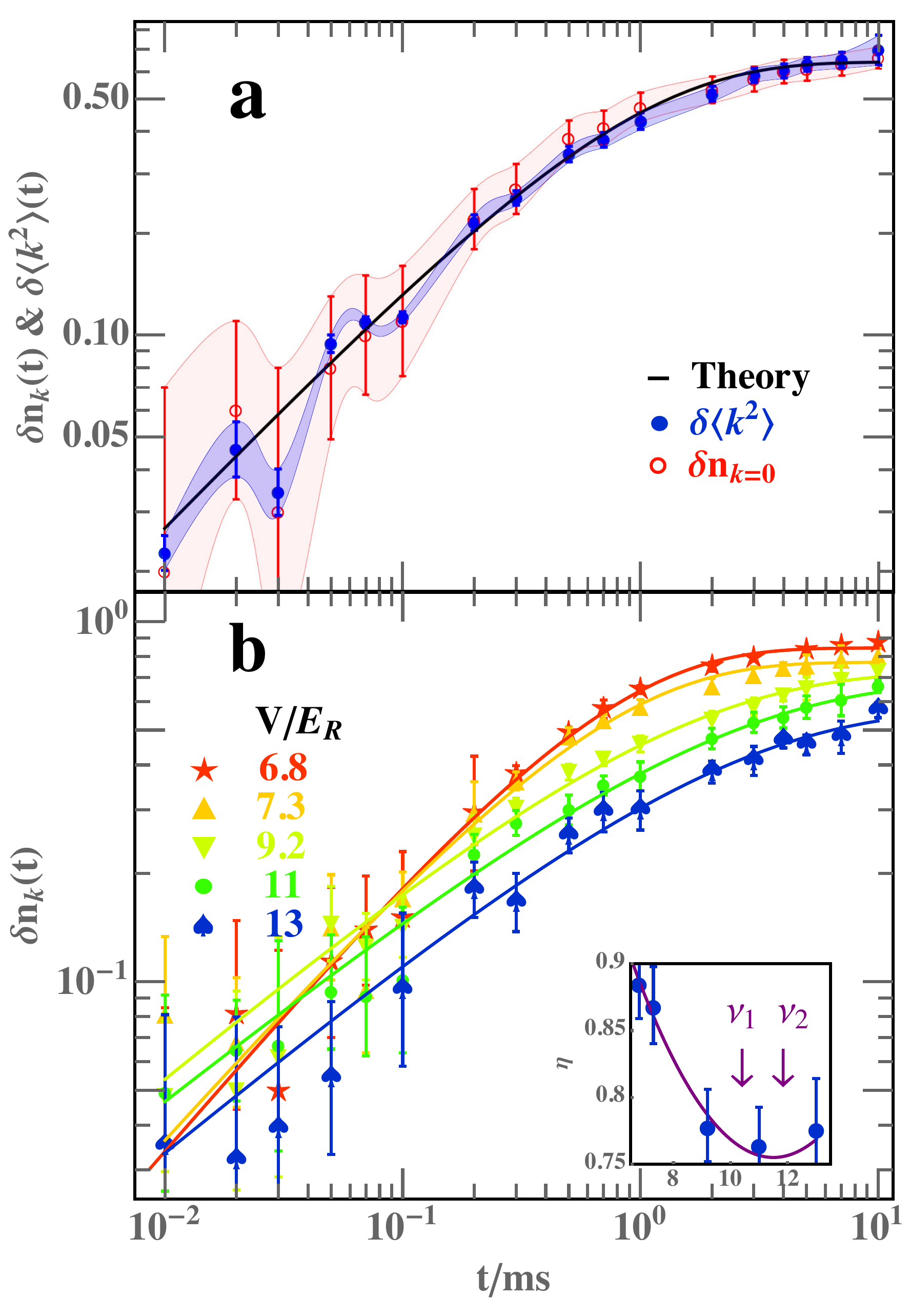}
	\caption{Reanalysis of Experimental Data on Dissipative Two-Dimensional Bose-Hubbard Model Reported in Ref. \cite{BHMExp} with Our Theory. (a) shows that two sets of data $\delta n_{{\bf k}=0}(t)$ and $\delta \langle k^2\rangle(t)$ perfectly coincide with each other by a properly chosen scaling factor. Solid line is fitting with our Eq. \ref{critical} with $\tau_0=0.7\text{ms}$. (b) Fit experimental data of $\delta n_{{\bf k}=0}(t)$ (scaled by $\delta n_{{\bf k}=0}(0))$ at different lattice depth with our Eq. \ref{critical}, which yields $\eta$ for different lattice depth shown in the inset. Two arrows label the critical value for superfluid-Mott insulator transition for filling number $\nu=1$ and $\nu=2$, respectively. $\tau_0$ for different curves are slightly different but all of them are of the order of $\sim 1\text{ms}$. All experimental data including error bars are taken from Ref. \cite{BHMExp}.} 
	\label{experiments}
\end{figure}

\begin{figure}[h]
\includegraphics[width=.9\columnwidth]{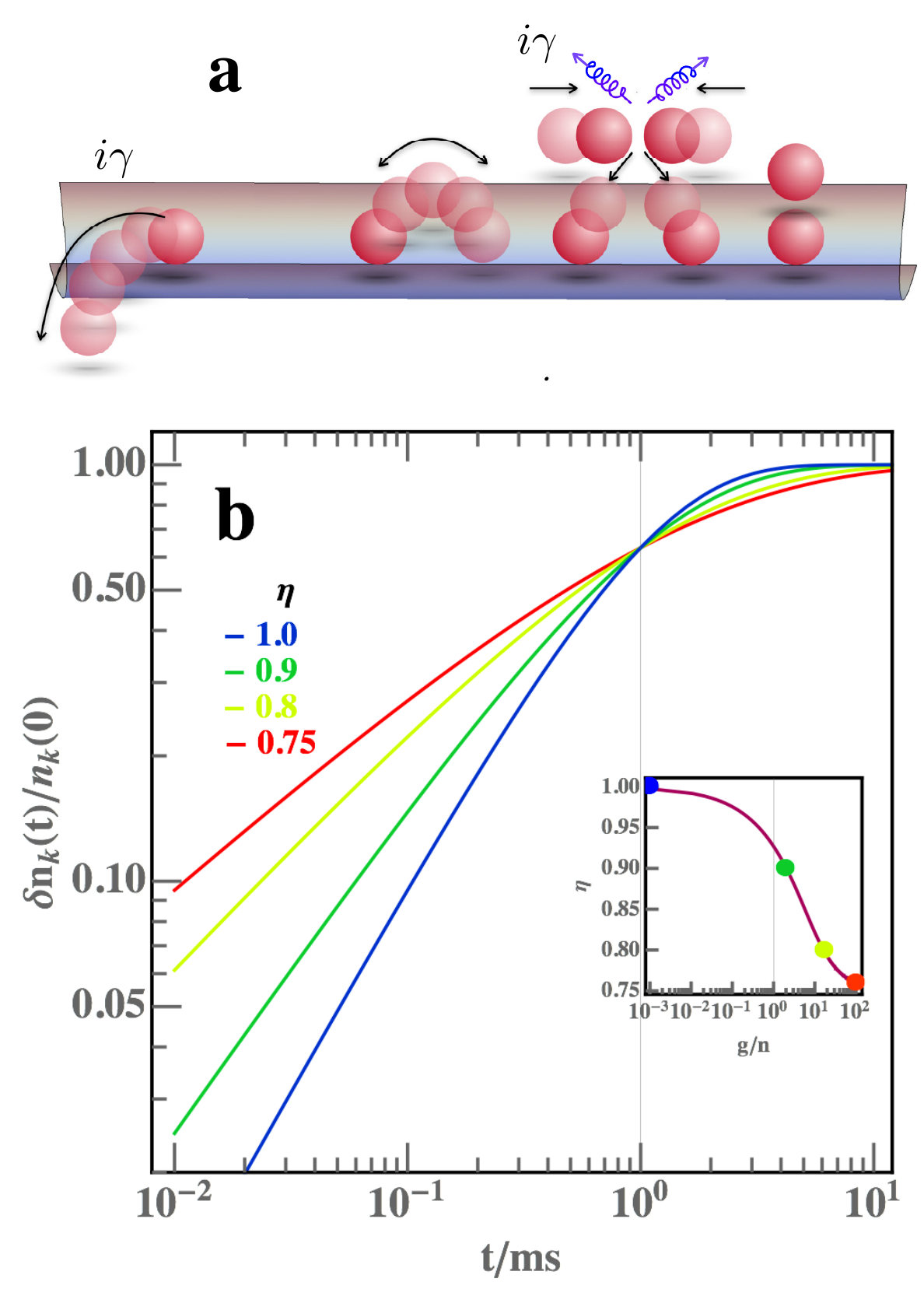}
\caption{Experimental Proposal of Dissipative Luttinger Liquid. (a) Schematic of a model of one dimensional bosons with hopping, interaction, one-body and two-body dissipations. (b) Predication of $\delta n_{{\bf k}}(t)$ for a one-dimensional Luttinger liquid from the weakly interacting limit (red line) to the Tonks limit (red line). Here $\tau_0=1.0\text{ms}$ is fixed for all plots. The inset shows how $\eta$ changes with the one-dimensional interaction parameter $g/n$. \label{oneD}}
\end{figure}

\textit{Width versus Height.} We introduce a quantity $\langle k^2\rangle=\int k^2n_{{\bf k}} /\int n_{{\bf k}}$ to characterize the width of the momentum distribution. We consider the situations that $f({\bf k},t)$ and $\mathcal{F}({\bf k},t)$ are independent of ${\bf k}$, as the examples discussed above. With Eq. \ref{sol_one}, it is straightforward to show that $\langle k^2\rangle$ is a constant independent of $t$. In this case, the one-body dissipation term does not change the width of the momentum distribution. With Eq. \ref{sol_two}, it can be shown that 
\begin{equation}
\langle k^2\rangle(t)-\langle k^2\rangle_0=(1-e^{-2\gamma \mathcal{F}(t)})\left(\langle k^2\rangle_\infty-\langle k^2\rangle_0\right), \label{k2}
\end{equation}
where $\langle k^2\rangle_0$ and $\langle k^2\rangle_\infty$ denote the width at $t=0$ and $t\rightarrow \infty$, respectively. By comparing Eq. \ref{k2} with Eq. \ref{sol_one_2} and Eq. \ref{sol_two_2}, we arrive at a significant prediction that, when $f({\bf k},t)$ and $\mathcal{F}({\bf k},t)$ are independent of ${\bf k}$, the change of momentum distribution for each momentum and the broadening of the width are governed by the same function.

\textbf{Experimental Predications.} 

Recently, a cold atom experiment has measured the real time dynamics of momentum distribution for the two-dimensional Bose-Hubbard model in the presence of weak dissipations \cite{BHMExp}. They have measured both the zero-momentum occupation as peak height and the width of the momentum distribution. Indeed, they find that in the weak lattice regime the width obeys a diffusion behavior, and the dynamics becomes slow for inter-mediate lattice depth \cite{BHMExp}. It is well known that there exists a superfluid to Mott insulator quantum phase transition at the intermediate lattice depth, and therefore, by our theory, we attribute this slowing down of dissipation induced dynamics to entering the quantum critical regime. This is a different explanation as original presented in Ref. \cite{BHMExp}. 

In the original paper of Ref. \cite{BHMExp}, the peak height and the width are presented as two separated measurements. However, our theory predicts that they should be governed by the same function Eq. \ref{critical} up to a coefficient. Guided by this theory, we re-plot the experimental data together as shown in Fig. \ref{experiments}(a). We find that, by choosing only one parameter as the relative coefficient, these two set of data points can overlap extremely well, in particular, for these data points with smaller error bars. This provides a very strong support to our theory.  

In Fig. \ref{experiments}(b), we show the critical exponent $\eta$ determined by fitting the real time dynamics of the momentum distribution reported in Ref. \cite{BHMExp} with our formula Eq. \ref{critical}, for the intermediate lattice depth. In the inset of Fig. \ref{experiments}(b) we show $\eta$ determined by this fitting as a function of lattice depth. Two arrows indicate quantum phase transition points for filling number $\nu=1$ and $\nu=2$. Since the maximum filling of this experiment is $\nu\approx 2.5$, due to the density inhomogeneity, these two critical values are relevant to this experiment. This fitting shows that $\eta$ approaches unity as the lattice depth becomes shallow and the system is away from the quantum critical regime, which is also consistent with diffusive behavior in the quasi-particle regime. This fitting also shows that $\eta$ is around $0.75$ in the quantum critical regime. To the best of our knowledge, such a critical exponent for the superfluid to Mott insulator transition has never been measured in cold atom system. It is also theoretically quite challenging to calculate this value because it is defined through the real time spectral function and the quantum Monte Carlo calculation faces the difficulty of the analytical continuation \cite{Sachdev02}. 

Hence, to provide a solid experimental benchmark of our theory, it is useful to experimentally study a system where $\eta$ can be calculated reliably in theory. To this end, we propose to experimental study a Luttinger liquid of either bosons or fermions with two-body dissipation, as shown in Fig. \ref{oneD}(a). The single-particle spectral function of a Luttinger liquid of one-dimensional bosons also behaves as Eq. \ref{spectral_Luttinger}, and $\eta$ is determined by the microscopic parameter $g/n$ \cite{Giamarchi,Lieb,Glazman1}, as shown in the inset of Fig. \ref{oneD}(b), where $g$ is the interaction strength and $n$ is the density. $\eta$ approaches $0.75$ in the strongly interacting Tonks limit and approaches $1$ in the weakly interacting limit. In Fig. \ref{oneD}(b), we show our predication of $\delta n_{{\bf k}}(t)$ as a function of $t$ for different interaction strengths. It shows that the difference between cases with different interaction strengths are significant enough that can be easily distinguished experimentally. As a side remark, it is interesting to note that, $\eta$ shown in the inset of Fig. \ref{experiments}(b) for the Bose-Hubbard model also varies from $~1$ to $~0.75$, as the system approaches the quantum critical points. 

\textbf{Outlook.} 

We believe our non-Hermitian linear response theory can be used for designing a new set of experiments, which are complementary to many existing experiments based on the Hermitian version of the linear response theory. As an example, we have successfully applied our theory to analyze a recent cold atom experiment on dissipative Bose-Hubbard model \cite{BHMExp}, and we show that the critical exponent of the real time spectral function can be extracted from this measurement. We suggest future experiments to carry out a systematically study of the temperature and interaction dependence of this critical exponent for two-dimensional Bose-Hubbard model, which is a quantity hard to calculate in theory. It can also be applied to possible other non-Fermi liquid states such as in the unitary Fermi gas and in the Fermi Hubbard model, where theoretical predications are also challenging. Thus, it can show the power of quantum simulation. Similar probe of spin dynamics, transport and other observables can also be studied both theoretically and experimentally. In addition, here we only considered a dc-perturbation in this work and further extension to non-Hermitian ac-perturbation can yield interesting results on non-Hermitian spectroscopy. We believe that our work opens a new avenue to probe Hermitian system with non-Hermitian perturbations.

\textit{Acknowledgment.} This work is supported by Beijing Outstanding Young Scientist Program (HZ), NSFC Grant No. 11734010 (HZ and YC), NSFC under Grant No. 11604225 (YC), MOST under Grant No. 2016YFA0301600 (HZ) and Beijing Natural Science Foundation (Z180013) (YC).

\begin{widetext}
\begin{appendix}
\setcounter{equation}{0}
\setcounter{subsection}{0}
\renewcommand{\theequation}{S\arabic{equation}}
\renewcommand{\thesubsection}{S\arabic{subsection}}

\centerline{\bf Supplementary Material}

\section{I. Derivation of the Linear Response}

As stated in the main text, we start from a hamiltonian with both dissipative terms and Langevin noise terms as follows,
\begin{equation}
\hat{H}=\hat{H}_0+\hat{H}_{\rm diss} ,\hspace{4ex}
\hat{H}_{\rm diss}=\sum_j\left(-i\gamma \hat{\cal O}_j^\dag \hat{\cal O}_j^{}+\hat{\cal O}_j^\dag\xi_j^{}+\xi^\dag_j \hat{\cal O}_j^{}\right),
\end{equation}
where $\xi$ and $\xi^\dag$ are the Langevin noise operators, $\hat{\xi}(t)=e^{i\hat{H}_0 t}\xi e^{-i\hat{H}_0 t}$, and they satisfy $\langle \hat{\xi}_{j}(t)\hat{\xi}_{\ell}^\dag(t')\rangle_{\rm noise}=2\gamma\delta_{j\ell}\delta(t-t')$, $\langle \hat{\xi}_j^\dag(t)\hat{\xi}^\dag_l(t')\rangle_{\rm noise}=0$, and $\langle\cdot\rangle_{\rm noise}$ stands for averaging over noise configurations. 

For any operator $\hat{W}$, the time evolution in this open system follows
\begin{equation}
\hat{W}_H(t)=e^{i\hat{H}^\dag t}\hat{W} e^{-i\hat{H} t}.
\end{equation}
By introducing $\hat{W}(t)=e^{i\hat{H}_0 t}\hat{W}e^{-i\hat{H}_0t}$ and evolution operator ${\cal U}(t)=e^{i\hat{H}_0 t}e^{-i\hat{H}t}$, we have
\begin{equation}
\hat{W}_H(t)={\cal U}^\dag (t)\hat{W}(t){\cal U}(t).\label{Heisenberg}
\end{equation}
In the interaction picture, the time evolution operator can be written in a time-ordered integral form as
\begin{equation}
{\cal U}(t)=T_t \left(\exp\left(-i \int_0^t \hat{H}_{\rm diss}(t')dt'\right)\right),
\end{equation}
where $T_t$ is time-ordered operator.
For small time $\gamma$, we can expand the time evolution operator as
\begin{equation}
{\cal U}(t)\approx1-i\int_0^t  \sum_j\left(-i\gamma \hat{\cal O}_j^\dag(t') \hat{\cal O}_j^{}(t')+\hat{\cal O}_j^\dag(t')\xi_j^{}(t')+\xi^\dag_j(t')\hat{\cal O}_j^{}(t')\right) dt'.\label{Eq:TimeOp}
\end{equation}
One can check that in linear response level (keeping ${\cal O}(\gamma)$ order)  and after noise average,
\begin{eqnarray}
{\cal U}^\dag (t) {\cal U}^{}(t)&=&1-\gamma\int_0^t\sum_j 2\hat{\cal O}_j^\dag(t')\hat{\cal O}_j(t') dt'+\sum_{j,\ell}\iint_0^t dt'dt'' \langle\hat{\cal O}_j^\dag(t')\hat{\cal O}_\ell(t'')\xi_j(t')\xi_{\ell}(t'')\rangle_{\rm noise}\nonumber\\
&=&1+2\gamma\int_0^t \sum_j(\hat{\cal O}_j^\dag(t')\hat{\cal O}_j(t')-\hat{\cal O}_j^\dag(t')\hat{\cal O}_j(t'))=1
\end{eqnarray}
It gives ${\cal U}^\dag (t) {\cal U}(t)=1$, and this means ${\cal U}(t)$ bears similarity as a unitary evolution operator in the Hermitian case. 
Applying Eq. (\ref{Eq:TimeOp}) to Eq. (\ref{Heisenberg}), we have
\begin{eqnarray}
\hat{W}_H(t)&=&\hat{W}(t)-\gamma\sum_j\int_0^t \{\hat{W}(t), \hat{\cal O}^{\dag}_j(t')\hat{O}_j^{}(t')\} dt'\nonumber\\
&&+\iint_0^t \left(\hat{\cal O}_j^\dag(t')\xi_j^{}(t')+\xi^\dag_j(t')\hat{\cal O}_j^{}(t')\right)\hat{W}(t)\left(\hat{\cal O}_j^\dag(t'')\xi_j^{}(t'')+\xi^\dag_j(t'')\hat{\cal O}_j^{}(t'')\right)dt'dt''.
\end{eqnarray}
First, we take the noise average, and with definition ${\cal \hat{W}}(t)\equiv \langle \hat{W}_H(t)\rangle_{\rm noise}$, we have
\begin{equation}
\hat{\cal W}(t)=\hat{W}(t)+\gamma\int_0^t \sum_j\left(2\hat{\cal O}_j^\dag(t')\hat{W}(t)\hat{\cal O}_j^{}(t')-\{\hat{W}(t), \hat{\cal O}^{\dag}_j(t')\hat{O}_j^{}(t')\}\right) dt'\label{OpEq}
\end{equation}
Then, we take ensemble average from two side of Eq. (\ref{OpEq}). With definition ${\cal W}(t)=\langle {\rm Tr}(\rho_0\hat{W}_H(t))\rangle_{\rm noise}$ and $\langle \hat{W}(t)\rangle=\langle \hat{W}\rangle$, we have
\begin{equation}
\delta {\cal W}(t)\equiv{\cal W}(t)-{\cal W}(0)=\gamma\int_0^t \sum_j\langle2\hat{\cal O}^\dag(t')\hat{W}(t)\hat{\cal O}_j^{}(t')-\{\hat{W}(t), \hat{\cal O}^{\dag}_j(t')\hat{O}_j^{}(t')\}\rangle dt'.\label{NHLRT}
\end{equation}
This is what we call non-Hermitian linear response theory in our paper. Notice that the Green's functions on the right hand side of the equation are Green's functions at unperturbed equilibrium state,.

\section{III. Momentum Distribution Response in one-body and two-body dissipative systems}

\subsection{One-Body Dissipation}

First, let us consider the single-body dissipation where $\hat{\cal O}_j=\hat{a}_j$ where $\hat{a}_{j}$ is the annihilation operator of the boson field at site $j$. Here we take the observable $\hat{W}$ to be momentum occupation number $\hat{n}_{\bf k}=\hat{a}_{\bf k}^\dag \hat{a}_{\bf k^{}}$.

By applying the non-Hermitian linear response theory, Eq. (\ref{NHLRT}), we have
\begin{eqnarray}
\delta n_{\bf k}(t)=\gamma \sum_i\left[\int_0^t dt' 2\langle \hat{a}_j^\dag(t')\hat{a}^\dag_{\bf k}(t) \hat{a}_{\bf k}^{}(t) \hat{a}_j^{}(t')\rangle-\int_0^t dt' \langle \hat{a}_j^\dag(t')\hat{a}_j^{}(t')\hat{a}^\dag_{\bf k}(t) \hat{a}_{\bf k}^{}(t)+\hat{a}^\dag_{\bf k}(t) \hat{a}_{\bf k}^{}(t)\hat{a}_j^\dag(t')\hat{a}_j^{}(t')\rangle\right].\label{Snk_one}
\end{eqnarray} 
Now we apply the Wick's theorem in above formula. By introducing the greater Green's function $G^>_{\bf k}(t_1-t_2)\equiv-i\langle a_{\bf k}^\dag(t_1) a_{\bf k}(t_2)\rangle$ and the lesser Green's function $G^<_{\bf k}(t_1-t_2)\equiv-i\langle a_{\bf k}^{}(t_1) a_{\bf k}^\dag(t_2)\rangle$, we find
\begin{eqnarray}
&&\sum_j\langle \hat{a}_j^\dag(t')\hat{a}^\dag_{\bf k}(t) \hat{a}_{\bf k}^{}(t) \hat{a}_j^{}(t')\rangle=\sum_j\sum_{{\bf k}_1,{\bf k}_2}\frac{1}{V^2}e^{-i{\bf k}_1\cdot {\bf r}_j}e^{i{\bf k}_2\cdot{\bf r}_j} \langle \hat{a}_{{\bf k}_1}^\dag(t')\hat{a}^\dag_{\bf k}(t) \hat{a}_{\bf k}^{}(t) \hat{a}_{{\bf k}_2}^{}(t')\rangle\nonumber\\
 &=&\sum_{{\bf k}_1,{\bf k}_2}\frac{1}{V}\delta({\bf k}_1-{\bf k}_2) \langle \hat{a}_{{\bf k}_1}^\dag(t')\hat{a}^\dag_{\bf k}(t) \hat{a}_{\bf k}^{}(t) \hat{a}_{{\bf k}_2}^{}(t')\rangle\approx\langle \contraction[1.2ex]{\ \ }{AAA\hspace{1.2cm}}{A}{}\contraction[2ex]{A}{B}{AAAA\ }{}a_{\bf k}^\dag(t') a_{\bf k}^\dag(t) a_{\bf k}^{}(t)a_{\bf k}^{}(t')\rangle+\sum_{\bf k'}\langle \contraction[1.2ex]{\ \ }{AAAA\hspace{0.25cm}}{A}{}\contraction[2ex]{A}{B}{AAAA\hspace{0.8cm}}{}a_{\bf k'}^\dag(t') a_{\bf k}^\dag(t) a_{\bf k}^{}(t)a_{\bf k'}^{}(t')\rangle\nonumber\\
 &=&\langle \hat{a}_{\bf k}^\dag(t') a_{\bf k}^{}(t)\rangle \langle\hat{a}_{\bf k}^\dag(t) a_{\bf k}^{}(t')\rangle+n_{\bf k} N =iG^>_{{\bf k}}(t'-t) iG^>_{{\bf k}}(t-t')+ n_{\bf k} N\label{Eq31}
\end{eqnarray}
where the $\approx$ is because we dropped all the terms with irreducible four point Green's functions.
Here we can prove that $iG^>_{\bf k}(t)\equiv\int n_B(\omega){\cal A}_{\bf k}(\omega)e^{i\omega t} d\omega$ and ${\cal A}_{\bf k}(\omega)$ is the single particle spectral function for bosonic field $a_{\bf k}$. Similarly, we find
\begin{eqnarray}
&&\sum_j\langle \hat{a}_j^\dag(t')\hat{a}_j^{}(t')\hat{a}^\dag_{\bf k}(t) \hat{a}_{\bf k}^{}(t)+\hat{a}^\dag_{\bf k}(t) \hat{a}_{\bf k}^{}(t)\hat{a}_j^\dag(t')\hat{a}_j^{}(t')\rangle\nonumber\\
&\approx&\sum_j\left(\langle \contraction[1.2ex]{\ \ }{AAA\hspace{1.2cm}}{A}{}\contraction[2ex]{A}{B}{AAAA\ }{}\hat{a}_j^\dag(t')\hat{a}_j^{}(t')\hat{a}^\dag_{\bf k}(t) \hat{a}_{\bf k}^{}(t)\rangle+\langle \contraction[1.2ex]{\!\!\!}{A\hspace{1cm}}{\ }\hat{a}_j^\dag(t')\hat{a}_j^{}(t')\contraction[1.2ex]{\!\!\!}{A\hspace{1cm}}{\ }\hat{a}^\dag_{\bf k}(t) \hat{a}_{\bf k}^{}(t)\rangle+\langle \contraction[1.2ex]{\ \ }{AAAA\hspace{0.25cm}}{A}{}\contraction[2ex]{A}{B}{AAAA\hspace{0.8cm}}{}\hat{a}^\dag_{\bf k}(t) \hat{a}_{\bf k}^{}(t)\hat{a}_j^\dag(t')\hat{a}_j^{}(t')\rangle+\langle \contraction[1.2ex]{\!\!\!}{A\hspace{1cm}}{\ }\hat{a}_{\bf k}^\dag(t')\hat{a}_{\bf k}^{}(t')\contraction[1.2ex]{\!\!\!}{A\hspace{1cm}}{\ }\hat{a}^\dag_{j}(t) \hat{a}_{j}^{}(t)\rangle\right)\nonumber\\
&=& iG^>_{\bf k}(t'-t) iG^{<}_{\bf k}(t'-t)+iG^>_{\bf k}(t-t') iG^{<}_{\bf k}(t-t')+2n_{\bf k}N.\label{Eq32}
\end{eqnarray}
Here $iG_{\bf k}^<(t)$ can be written in terms of spectral function as $\int (1+n_B(\omega)){\cal A}_{\bf k}(\omega)e^{-i\omega t}d\omega$. Combining Eq. (\ref{Eq31}) and Eq. (\ref{Eq32}), we obtain
\begin{eqnarray}
\delta n_{\bf k}(t)
&\approx&\gamma \int_0^t \left[iG^>_{\bf k}(t'-t)(iG_{\bf k}^>(t-t')-iG_{\bf k}^{<}(t'-t))+iG_{\bf k}^{>}(t-t')(iG^>_{\bf k}(t'-t)-iG_{\bf k}^<(t-t'))\right]dt'\nonumber\\
&=&-\gamma \int_0^t \left[(\int d\omega n_B(\omega) {\cal A}_{\bf k}(\omega) e^{i\omega (t'-t)}) g_{\bf k}(t-t')+g_{\bf k}(t'-t)(\int d\omega n_B(\omega) {\cal A}_{\bf k}(\omega) e^{-i\omega (t'-t)})\right]\nonumber\\
&\approx&-2\gamma n_{\bf k}(0) \int_0^t g_{\bf k}(t'-t) g_{\bf k}(t-t') dt'=-2\gamma n_{\bf k}(0) \int_0^t g_{\bf k}(t') g_{\bf k}(-t') dt',\label{OneBody}
\end{eqnarray}
where $g_{\bf k}(t)\equiv\int d\omega {\cal A}_{\bf k}(\omega)e^{i\omega t}$.
In the last equation we used an approximation that $\int d\omega n_B(\omega){\cal A}(\omega)e^{i\omega t}\approx n_{\bf k}(0) g_{\bf k}(t)$. This approximation usually works well when the single particle spectrum is peaked at the frequency $\omega=\epsilon_{\bf k}$, as in cases discussed in the main text. As we can change $t-t'$ as $t'$, so the two formula in the last line are equivalent. Eq. (\ref{OneBody}) is the equation we obtained in the main text for single-particle dissipation.
\newline

\subsection{Two-Body Dissipation}

Second, we discuss the situation for two-body dissipation where $\hat{\cal O}_j=\hat{n}_{j}$. The observable is still the momentum distribution $\hat{n}_{\bf k}$. According to our formula, we have
\begin{eqnarray}
\delta n_{\bf k}(t)&&=\gamma \sum_j\left[\int_0^t dt' 2\langle \hat{n}_j^\dag(t')\hat{a}^\dag_{\bf k}(t) \hat{a}_{\bf k}^{}(t) \hat{n}_j^{}(t')\rangle-\int_0^t dt' \langle \hat{n}_j^\dag(t')\hat{n}_j^{}(t')\hat{a}^\dag_{\bf k}(t) \hat{a}_{\bf k}^{}(t)+\hat{a}^\dag_{\bf k}(t) \hat{a}_{\bf k}^{}(t)\hat{n}_j^\dag(t')\hat{n}_j^{}(t')\rangle\right].\label{Snk_two}
\end{eqnarray}
Again, we apply the Wick's theorem to Eq. (\ref{Snk_two}). Notice that there are always contraction between two equal-time field operators, which always yields average density $\langle n_j\rangle=\bar{n}$. We find that 
\begin{eqnarray}
\langle \hat{n}_j^\dag(t')\hat{a}^\dag_{\bf k}(t) \hat{a}_{\bf k}^{}(t) \hat{n}_j^{}(t')\rangle&\approx&\langle \contraction[1.2ex]{}{\!\!AA\hspace{4ex}}{}\hat{a}_j^\dag(t') \hat{a}_j(t') \contraction[1.2ex]{\ \ }{AAAA}{A}{}\contraction[2ex]{A}{B}{AAAA\hspace{0.8cm}}{}\hat{a}^\dag_{\bf k}(t) \hat{a}_{\bf k}(t)^{} \hat{a}_j^{\dag}(t')\hat{a}_j^{}(t')\rangle+\langle \contraction[1.2ex]{\ \ }{AAA\hspace{0.4cm}}{A}{}\contraction[2ex]{A}{B}{AAAA\hspace{1cm}}{}\hat{a}_j^\dag(t')\hat{a}_j^{}(t')\hat{a}^\dag_{\bf k}(t) \hat{a}_{\bf k}^{}(t)\contraction[1.2ex]{}{\!\!AA\hspace{4ex}}{}\hat{a}_j^\dag(t') \hat{a}_j(t')\rangle+\nonumber\\
&&\langle \contraction[1.2ex]{\ \ }{AAA\hspace{0.4cm}}{A}{}\contraction[2ex]{A}{B}{AAAA\hspace{2.5cm} }{}\hat{a}_j^\dag(t')\hat{a}_j^{}(t') \contraction[1.2ex]{\ \ }{AAA\hspace{0.4cm}}{A}{}\hat{a}_{\bf k}^\dag(t)\hat{a}_{\bf k}^{}(t)\hat{a}_j^\dag(t') \hat{a}_j^{}(t')\rangle+\langle \contraction[1.2ex]{\!\!\!\!\!\!\!\!}{AAA\hspace{3cm}}{A}{}\contraction[2ex]{A}{B}{AAAA\hspace{1.2cm} }{}\hat{a}_j^\dag(t')\contraction[2.6ex]{\!\!\!\!\!}{AA\hspace{2.8cm}}{A}{}\hat{a}_j^{}(t')\hat{a}_{\bf k}^\dag(t)\hat{a}_{\bf k}^{}(t)\hat{a}_j^\dag(t') \hat{a}_j^{}(t')\rangle\rangle+\bar{n}^2 n_{\bf k}\nonumber\\
&=&\bar{n}^{\frac{}{}}(i G^>_{{\bf k},j}(t-t') iG_{{\bf k},j}^<(t-t'))+\bar{n}^{\frac{}{}}(i G^>_{j,{\bf k}}(t'-t) iG_{j,{\bf k}}^<(t'-t))+\bar{n}^2 n_{\bf k}+\nonumber\\
&&\bar{n}^{\frac{}{}}(i G^<_{j,{\bf k}}(t'-t) iG_{{\bf k},j}^<(t-t'))+(1+\bar{n})(i G^>_{j,{\bf k}}(t'-t) iG_{{\bf k},j}^>(t-t')).\label{S1}
\end{eqnarray}
Following the decomposition in Eq.~(\ref{S1}), we can further find 
\begin{eqnarray}
\langle \hat{n}_j^\dag(t')\hat{n}_j^{}(t')\hat{a}^\dag_{\bf k}(t) \hat{a}_{\bf k}^{}(t)\rangle&\approx&(2\bar{n}+1+2\bar{n})(i G^>_{j,{\bf k}}(t'-t) iG_{j,{\bf k}}^<(t'-t))+\bar{n}^2n_{\bf k},\label{S2}\\
\langle \hat{a}^\dag_{\bf k}(t) \hat{a}_{\bf k}^{}(t)\hat{n}_j^\dag(t')\hat{n}_j^{}(t')\rangle&\approx&(2\bar{n}+1+2\bar{n})(i G^>_{{\bf k},j}(t-t') iG_{{\bf k},j}^<(t-t'))+\bar{n}^2n_{\bf k}.\label{S3}
\end{eqnarray}
By inserting Eq. (\ref{S1}), Eq. (\ref{S2}) and Eq. (\ref{S3}) to Eq. (\ref{Snk_two}), we obtain
\begin{eqnarray}
\delta n_{\bf k}(t)&\approx&\gamma\sum_{j}\left[\int_0^t \left((2+2\bar{n})(i G^>_{j,{\bf k}}(t'-t) iG_{{\bf k},j}^>(t-t'))+2\bar{n}(i G^<_{j,{\bf k}}(t'-t) iG_{{\bf k},j}^<(t-t'))\right.\right.\nonumber\\
&&\left.\left.-(2\bar{n}+1)(i G^>_{{\bf k},j}(t-t') iG_{{\bf k},j}^<(t-t'))-(2\bar{n}+1)(i G^>_{j,{\bf k}}(t'-t) iG_{j,{\bf k}}^<(t'-t))\right)dt'\right]\nonumber\\
&=&\gamma\left[\int_0^t \left((2+2\bar{n})(i G^>_{{\bf k}}(t'-t) iG_{{\bf k}}^>(t-t'))+2\bar{n}(i G^<_{{\bf k}}(t'-t) iG_{{\bf k}}^<(t-t'))\right.\right.\nonumber\\
&&\left.\left.-\frac{}{}(2\bar{n}+1)(i G^>_{{\bf k}}(t-t') iG_{{\bf k}}^<(t-t'))-(2\bar{n}+1)(i G^>_{{\bf k}}(t'-t) iG_{{\bf k}}^<(t'-t))\right)dt'\right]\nonumber\\
&=&\gamma(2\bar{n}+1)\int_0^t (iG_{\bf k}^<(t-t')-iG^>_{\bf k}(t'-t))(iG_{\bf k}^>(t-t')-iG_{\bf k}^<(t'-t))dt'\nonumber\\
&&+\gamma\int_0^t\left[i G^>_{{\bf k}}(t'-t) iG_{{\bf k}}^>(t-t')-i G^<_{{\bf k}}(t'-t) iG_{{\bf k}}^<(t-t')\right]dt'
\end{eqnarray}
By introducing approximation $\int d\omega n_B(\omega){\cal A}(\omega)e^{i\omega t}\approx n_{\bf k}(0) g_{\bf k}(t)$, we obtain
\begin{eqnarray}
\delta n_{\bf k}(t)&=&\gamma\left[(2\bar{n}+1)  +(n_{\bf k}^2(0)-(n_{\bf k}(0)+1)^2)\right]\int_0^t g_{\bf k}(t'-t)g_{\bf k}(t-t')dt'\nonumber\\
&=&-2\gamma(n_{\bf k}(0)-\bar{n})\int_0^t g_{\bf k}(t')g_{\bf k}(-t')dt'.
\end{eqnarray}
\end{appendix}
\end{widetext}

\end{document}